\documentclass[letterpaper, 10 pt, conference]{ieeeconf}  

\IEEEoverridecommandlockouts                              

\usepackage{graphicx} 
\usepackage{subfigure, color,url}
\usepackage{amsmath} 

\begin{document}

\title{\LARGE \bf
Toward QoE-driven Dynamic Control Scheme Switching for Time-Delayed Teleoperation Systems: A Dedicated Case Study
}

\author{Xiao Xu, Qian Liu and Eckehard Steinbach
\thanks{This work was supported, in part, by an Alexander von Humboldt Foundation Fellowship for postdoctoral researchers for Q. Liu.}
\thanks{X. Xu and E. Steinbach are with the Chair of Media Technology, Technical University of Munich, Germany. Emails:
        {\tt\small \{xiao.xu, eckehard.steinbach\}@tum.de}}%
\thanks{Qian Liu is with the Chair of Media Technology and the Chair of Communication Networks, Technical University of Munich, Germany. She is also with the Dept. of Computer Science and Technology, Dalian University of Technology, Dalian, China. Email:
        {\tt\small qianliu@dlut.edu.cn}}%
}

\maketitle

\begin{abstract}

Networked teleoperation with haptic feedback is a prime example for the emerging Tactile Internet,
which requires a careful orchestration of haptic communication and control. One major challenge in this context is how to maximize the user's quality-of-experience (QoE) while ensuring at the same time the stability
of the global control loop in the presence of communication delay. In this paper, we propose a dynamic control scheme switching strategy for teleoperation systems, which maximizes the QoE for time-varying communication delay. In order to validate the feasibility of the proposed approach, we perform a dedicated case study for a virtual teleoperation environment consisting of a one-dimensional spring-damper system, and conduct extensive subjective tests under various delay conditions for two control schemes : (1) teleoperation with the time-domain
passivity approach (TDPA), which is highly delay-sensitive but
supports highly dynamic interaction between the operator and
a potentially quickly changing remote environment; (2) model-mediated
teleoperation (MMT), which is tolerable to relatively
larger communication delays, but unsuitable for quickly changing, highly dynamic
remote environments. For both schemes, we use recently proposed extensions, which incorporate perceptual data reduction to reduce the required packet rate between the operator and the teleoperator. One key contribution of this paper lies in the exploration of the intrinsic relationship among QoE, communication delay and the control schemes which provides a fundamental guidance, not only to this research, but also to the future joint optimization of communication and control for time-delayed teleoperation systems.

\end{abstract}

\section{INTRODUCTION}

Teleoperation systems with haptic feedback allow a human user to immerse into a distant or
inaccessible environment to perform complex tasks. A typical teleoperation system comprises a slave and a master device, which exchange haptic information (forces, torques,
position, velocity), video signals, and audio signals over a communication network \cite{ferrell1967supervisory}.
In particular, the communication of haptic information (position/velocity and force/torque signals) imposes strong demands on the communication network as it closes a global control loop between the operator and the remote robot. As a result, the system stability is highly sensitive to communication delay \cite{lawrence1993stability}. In addition, high-fidelity teleoperation requires a high sampling rate for the haptic signals of 1 kHz or even higher \cite{colgate1997passivity} to ensure a high quality interaction and system stability. Teleoperation systems, hence, require 1000 or more data packets per second to be transmitted between the master and the slave device. For Internet-based communication, such high packet rates are hard to maintain.

\begin{figure}[t]
  \centering
    \includegraphics[width=0.35\textwidth]{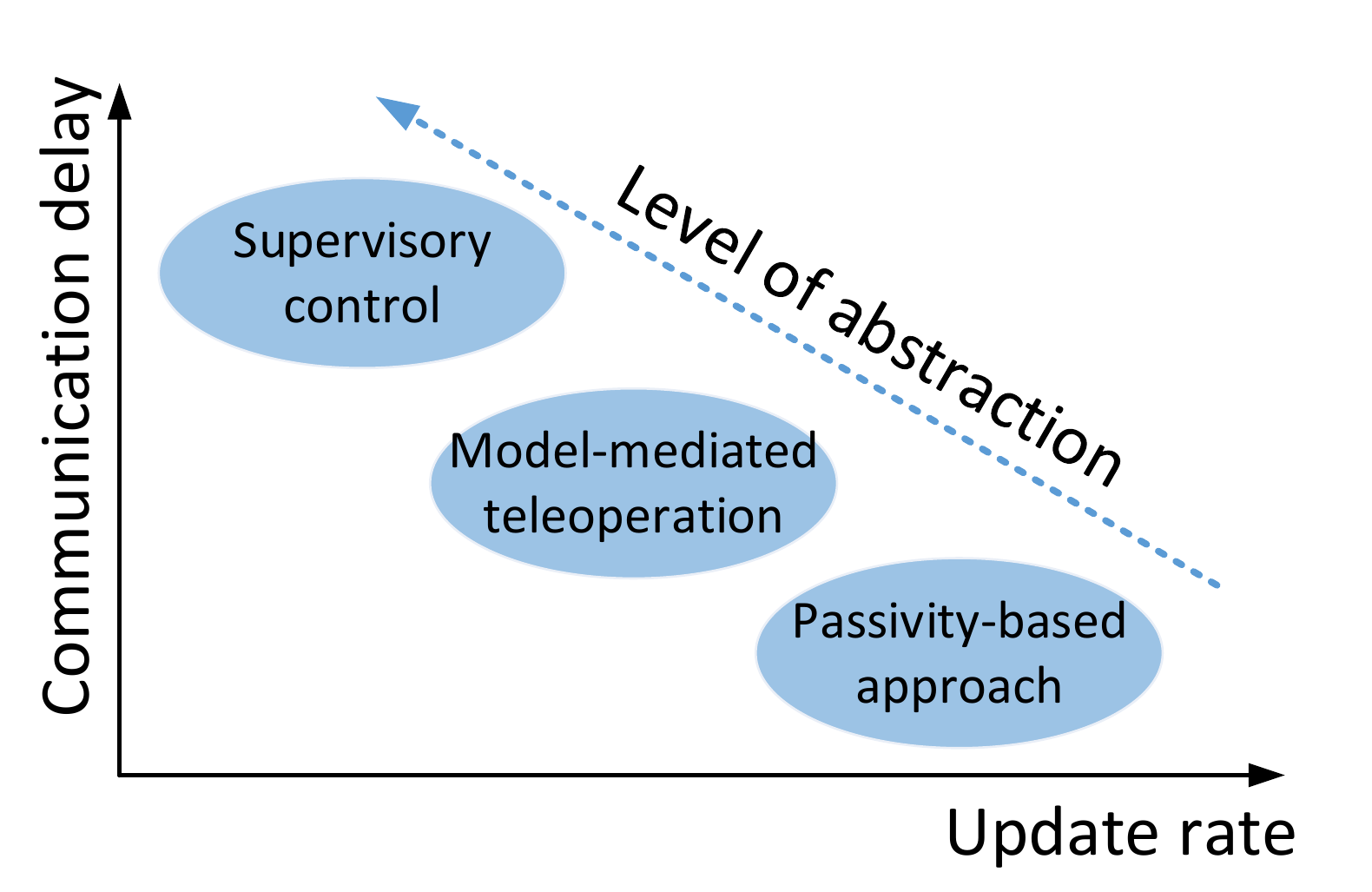}
\vspace{-0.1 in}
\caption{The level of abstraction and data complexity in teleoperation systems with update rates and robustness to delays (adapted from \cite{mitra2008model}).}

 \label{Fig::abs_vs_delay}
\end{figure}

State-of-the-art solutions that address the aforementioned teleoperation challenges (sensitivity to delay and high packet rate) focus on combining different stability-ensuring control schemes with haptic packet rate reduction methods such as \cite{Xu2016_IEEEToH, Xu2014_TIM}.

On the other hand, since the communication delay can range from a few milliseconds up to several hundred milliseconds, teleoperation systems require a control scheme which stabilizes the teleoperation system in the presence of end-to-end communication delays that are larger than a couple of milliseconds. Fig. \ref{Fig::abs_vs_delay} presents a qualitative analysis of the trade-off between the level of communication delay and the level of abstraction in control schemes for teleoperation \cite{mitra2008model}. We can observe from Fig. \ref{Fig::abs_vs_delay} that passivity-based control, e.g. the time-domain passivity approach (TDPA) described in \cite{Ryu2007, Ryu2010, Artigas2011}, is suitable for short-distance (low-latency) teleoperation with dynamic scenes and a high level of interaction between the master and the remote environment (i.e. high update rate); the model-mediated teleoperation (MMT) \cite{mitra2008model, Hannaford1989, Willaert2012} approach is able to deal with relatively larger communication delays (i.e. for medium or long distance application scenarios), but is unsuitable for quickly changing environments. Teleoperation for very large delay is typically performed using supervisory control and will not be further considered in this paper.

Although all control schemes, in theory,  are able to ensure system stability for arbitrary delays, different control schemes have different delay tolerance, introduce different artifacts which degrade the user experience, and require different amounts of resources from the communication network.  In short, the teleoperation performance varies with respect to different communication and control schemes. For all considered control schemes, the user's quality-of-experience (QoE) degrades for increasing communication delay. In literature, the impact of communication impairments on the teleoperation performance has been studied objectively and subjectively using the concept of transparency \cite{lawrence1993stability, hannaford1989stability, hashtrudi2000analysis} and perceptual transparency \cite{Hirche2007b, Hirche2012}. In addition, the authors of \cite{arcara2002control} compared 10 different control schemes in terms of stability region, position and force tracking performance, displayed impedance, position drift, etc. However, the control schemes in \cite{arcara2002control} were developed before 2002 and the authors did not consider the impact of haptic data reduction schemes.

Motivated by the above analysis, we propose in this paper a novel solution for time-delayed teleoperation systems  based on the state-of-the-art control schemes with perceptual haptic data reduction.  The proposed solution maximizes the user's QoE by dynamic switching among different control schemes with respect to different round-trip communication latencies. In order to validate the feasibility of the proposed strategy, we perform a dedicated case study for a virtual teleoperation environment consisting of a one-dimensional spring-damper system.

The remainder of this paper is organized as follows. In Section II, we will give a brief introduction to the considered control schemes for time-delayed teleoperation. In Section III, we will describe  the proposed dynamic control scheme switching strategy, which is then validated by a spring-damper experimental setup in Section IV. This case study also demonstrates the tight relationship among QoE, communciation delay and the used control schemes. Finally, Section V concludes this paper with a summary.

\section{Characteristics of Time-delayed Teleoperation with Different Control Schemes}

In order to address the two major communication-related challenges (i.e. time delay and high packet rate) of teleoperation systems, various schemes have been developed by combining different stability-ensuring control approaches with haptic data reduction algorithms. The most representative solutions from the literature are the combination of the TDPA from \cite{Ryu2010} with the perceptual deadband-based (PD) haptic data reduction solution from \cite{hinterseer2008perception}, denoted as TDPA+PD \cite{Xu2016_IEEEToH} in the following, and the combination of the MMT method from \cite{Hannaford1989, Willaert2012} with the haptic data reduction solution from \cite{hinterseer2008perception}, denoted as MMT+PD \cite{Xu2014_TIM} in the following. We will briefly introduce these two approaches in the following two sections.

\subsection{TDPA with Perceptual Haptic Data Reduction}

\begin{figure}[t]
  \centering
    \includegraphics[width=0.45\textwidth]{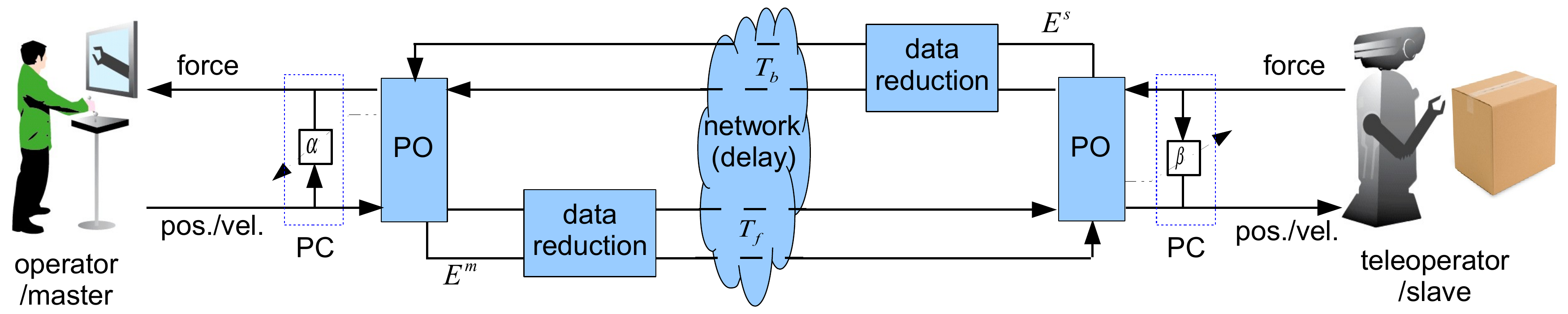} \\
\footnotesize (a) TDPA+PD, adopted from \cite{Xu2016_IEEEToH}
\vspace{0.05 in}
    \vfil
    \includegraphics[width=0.45\textwidth]{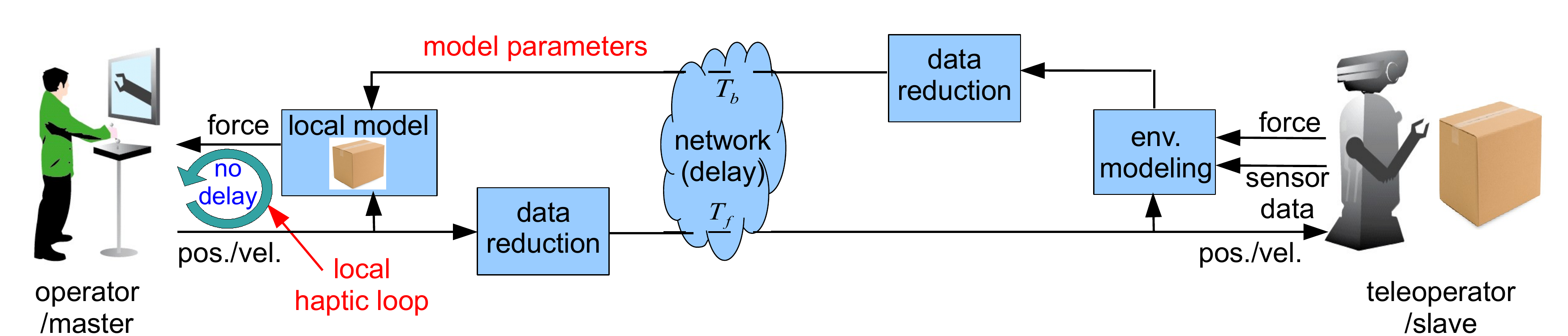} \\
  \footnotesize  (b) MMT+PD, adopted from \cite{Xu2014_TIM}
\caption{Overview of different control schemes with haptic data reduction. }
 \label{Fig::controlScheme}
\end{figure}

The TDPA \cite{Ryu2007, Ryu2010, Artigas2011} is a typical passivity-based control scheme for time-delayed teleoperation. The stability arguments are based on the passivity concept, which characterizes the energy exchange over a two-port network and provides a sufficient condition for the input/output stability. The stability of TDPA-based teleoperation systems is guaranteed in the presence of arbitrary communication delays with the help of passivity observers (PO) and passivity controllers (PC). The PO computes the current system energy. The PC adaptively adjusts the customized dampers $\alpha$ and $\beta$ to dissipate energy and thus guarantees the passivity of the system. In \cite{Xu2016_IEEEToH}, perceptual data reduction is incorporated into the TDPA approach. The resulting scheme is called TDPA+PD in the following.
The haptic data reduction blocks are placed after the POs to irregularly downsample the transmission of haptic packets using perceptual thresholds (see Fig.~\ref{Fig::controlScheme} (a)).


The TDPA is a conservative control design. With increasing delay, it leads to larger distortions in the displayed environment properties (e.g., hard objects are displayed softer than they actually are). In addition, the TDPA+PD method can lead to sudden force changes when the PCs are activated to dissipate the system output energy. This effect becomes stronger with increasing communication delays \cite{ Ryu2010, Xu2016_IEEEToH}.

Therefore, the TDPA+PD approach is suitable for short-distance teleoperation applications which may operate at the edge of the communication network in order to meet the requirement of frequent haptic updating between the master and the slave. Thus, it can deal with a high level of dynamics of the objects (motion, deformation, etc.) and interaction patterns.

\subsection{MMT with Perceptual Haptic Data Reduction}
\label{subsec::MMT}

One major issue of passivity-based control schemes is that the system passivity and transparency are conflicting objectives. This means that the system gains stability at the cost of degraded transparency \cite{lawrence1993stability}. In order to overcome this issue, model-mediated teleoperation (MMT) was proposed to guarantee both stability and transparency in the presence of communication delays \cite{Hannaford1989, Willaert2012}. In the MMT approach, a local object model is employed on the master side to approximate the slave environment. The model parameters describing the object in the slave environment are continuously estimated in real time and transmitted back to the master whenever the slave obtains a new model. On the master side, the local model is reconstructed or updated on the basis of the received model parameters, and the haptic feedback is computed on the basis of the local model without noticeable delay. If the estimated model is an accurate approximation of the remote environment, both stable and transparent teleoperation can be achieved \cite{mitra2008model, Passenberg2010}.  In \cite{Xu2014_TIM}, perceptual data reduction is incorporated into the MMT approach. The resulting scheme is called MMT+PD in the remainder of this paper. The data reduction scheme is used to irregularly downsample the velocity signals in the forward channel and the model parameters in the backward channel, using perceptual thresholds (see Fig.~\ref{Fig::controlScheme} (b)). For the model parameters, these thresholds determine whether a model update leads to a perceivable difference in the displayed signal. If not, the model change does not need to be transmitted.

Using the MMT approach, hard objects will not be displayed softer with increasing delay as for the TDPA. Therefore, the MMT approach has better teleoperation quality than the TDPA when the delay is relatively large. However, keeping the local model consistent with the environment at the remote side is challenging for dynamic scenes. This way, the MMT approach is favorable for medium/long distance teleoperation applications and scenarios which are characterized by a low level of scene dynamics.

Although the two approaches address different scenarios, it holds for both that the lower the end-to-end delay, the better the system transparency and hence the QoE.

\section{Proposed QoE-driven Dynamic Control Scheme Swithing Strategy}

Based on the analysis of Section II, we can conclude that different control and communication approaches introduce different types of artifacts into the system. Their performance varies between tasks (e.g. free space versus contact, soft objects versus rigid surface, etc.), and also differs in the robustness towards different network delays, which is the focus of this paper.

\begin{figure}[!t]
	\centering
	\includegraphics[height=2.8cm,width=0.34\textwidth]{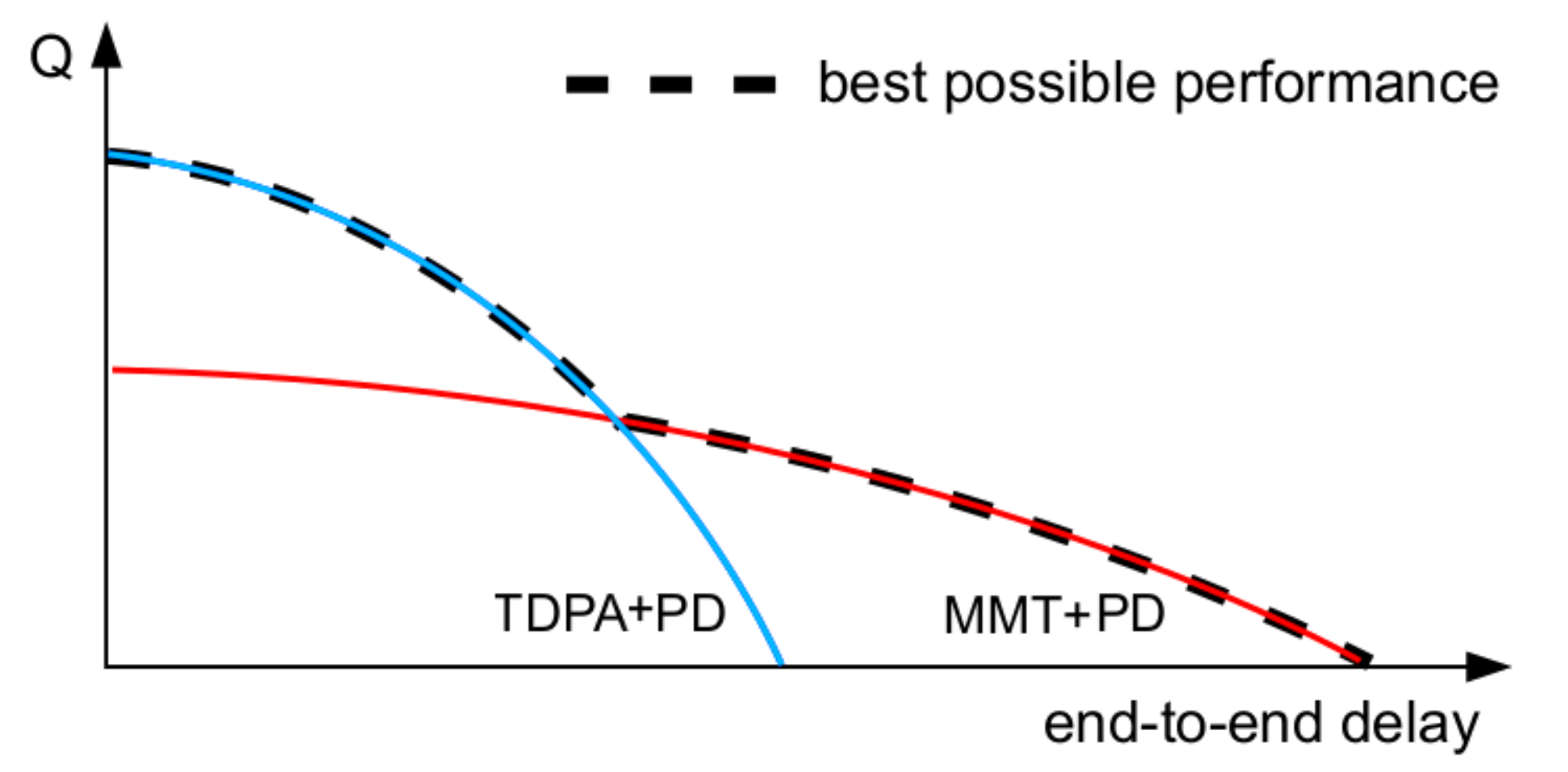}
\vspace{-0.1 in}
	\caption{Hypothetical performance metric as a function of the round-trip delay and control schemes.} \vspace{-0.15in}
	 \label{ch5:Fig:JointOpt}
\end{figure}

For example, the performance of the TDPA+PD method is mainly influenced by communication delay. The larger the delay, the softer the displayed impedance (stiffness), and the stronger the distortion introduced in the force signals. On the other hand, the performance of the MMT+PD method depends strongly on the model accuracy, which is barely affected by the communication delay for services with low object dynamics. Hence, the TDPA+PD and the MMT+PD methods are the best choice for different delay ranges. In addition, these two methods transmit different types of data in the backward channel, which can lead to very different traffic characteristics over the communication network.

Therefore, the teleoperation system needs to adaptively switch between different control schemes according to the current communication delay in order to achieve the best possible performance. For this, we propose the following joint optimization problem for the currently to be used control approach $\gamma_c $
\begin{equation}
\gamma_c = \mathrm{arg}\max\limits_{\gamma_i \in \Gamma}  Q(\tau, \gamma_i)
\end{equation}
where $\tau$ represents the round-trip communication delay, $\gamma_i$ represents the $i$-th control scheme from the set of available control schemes $\Gamma$, and $Q(\tau, \gamma_i)$ represents the teleoperation performance (e.g., expressed in terms of the user's QoE) which is a function of both communication delay and the control scheme. In the next section, we will derive $Q(\tau, \gamma_i)$ for the TDPA+PD and MMT+PD methods through subjective tests.

Hypothetically, the QoE as a function of the communication delay for the two control schemes (TDPA+PD and MMT+PD) can be illustrated as shown in Fig.~\ref{ch5:Fig:JointOpt} (which will be verified with an experimental case study in the next section). We should point out that Fig.~\ref{ch5:Fig:JointOpt} reveals the fundamental trade-off among QoE, communication delay and control approaches. This way, the optimal solution of the proposed optimization problem can be obtained through dynamic switching between control schemes based on the current delay conditions.

\section{Feasibility Validation of the Proposed Design: A Case Study}\label{sec:traffic}

\begin{figure}[t]
	\centering
	\includegraphics[width=0.4\textwidth]{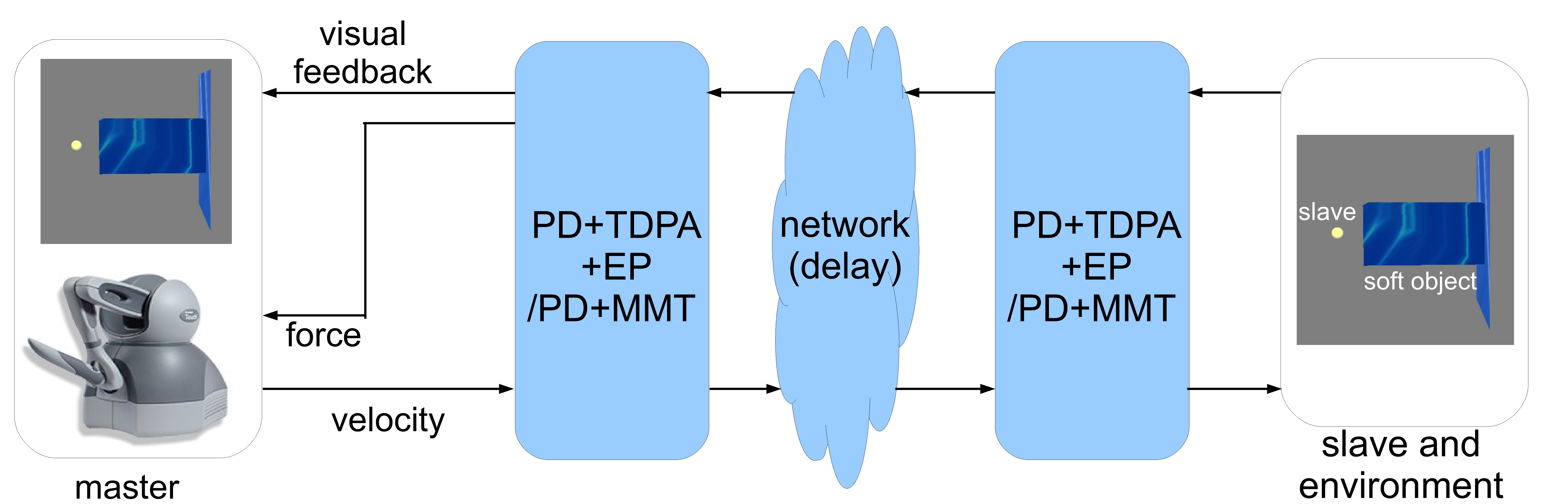}
	\caption{Experimental setup. The communication network, the slave (represented by a haptic interaction point), and the virtual environment are simulated on a PC using the CHAI3D library.} \vspace{-0.1in}
	\label{Fig::expSetup}
\end{figure}

In this section, we use a simple one-dimentional spring-damper setup (as shown in Fig. \ref{Fig::expSetup})  to validate the feasibility of the proposed switching solution. In particular, we evaluate and compare the performance of the two previously discussed control methods in terms of user preference and the generated haptic data traffic. The user's QoE for both control schemes in the presence of difference communication delays is evaluated using subjective tests.

\subsection{Experimental Setup}

The experiment was conducted in a virtual environment (VE) developed based on the Chai3D library (www.chai3d.org). The Phantom Omni haptic device was used as the master, while the slave in the VE was designed as a single haptic interaction point (HIP) with negligible mass. The communication network with adjustable delay was simulated in a local PC.
The VE contained a 1D non-linear soft object, whose interaction force $f_e$ is computed using the Hunt-Crossley model \cite{Hunt1975}
\begin{equation}
\label{Eq::HuntCrossley}
f_e=\left\{
\begin{matrix}
Kx^n+Bx^n\dot{x} + \Delta f & x \ge 0 \\
0 & x < 0
\end{matrix}\right.
\end{equation}
where $\Delta f$ is Gaussian distributed measurement noise with a mean of 0 $N$ and a standard deviation of 0.1 $N$. $x$ denotes the penetration (compressed displacement).
Corresponding parameters were set as: $K=200$ N/m, $n=1.5$, and $B=0.5$ N/ms. For the PD+MMT method, a simple linear  spring model ($\hat{f}_e = \hat{K}x$) was employed to approximate the environment. This leads to model mismatch and frequent changes in the estimated model stiffness $\hat{K}$ during interaction. The passivity-based model update scheme, introduced in \cite{xu2015passivity}, was applied to ensure stable model updates on the master side. All experiments were conducted on a PC with an Intel Core i7 CPU and 8~GB memory. The whole experimental setup is illustrated in Fig.~\ref{Fig::expSetup}.

The round-trip delays were set to 0 ms, 10 ms, 25 ms, 50 ms, 100 ms, and 200 ms, respectively. For each delay, the subjects tested three systems: the TDPA+PD method, the MMT+PD method, and the 0-delay reference without using any control and data reduction schemes. The reference scenario was shown to the subjects before the real experiments. The original environment impedance was displayed and the best performance (uncompressed, non-delayed) for this setup was provided.

The subjects interacted with the virtual object by pressing its surface and slowly varying the applied force.
The subjects were asked to give a rating by comparing the interaction quality between each control scheme and the reference scenario. They were asked to take all perceivable distortions (e.g. force vibrations, force jumps, perceived impedance variations, etc.) into account when evaluating the interaction quality. The rating scheme was based on Table~\ref{Tab::subjectiveRating}. The reference, designated level 5, was considered as the best performance. The reference can be recalled by the subjects at any time during the experiment.
Each delay case was repeated four times. The order of the tested delay as well as the order of the tested control methods were randomly selected.

There were 12 participants, i.e. subjects (right handed and ranging from the age of 25 to 33), in the experiment. The participants were asked to wear a headset with active noise cancellation to protect them from the ambient noise. During the experiment, they were first provided with a training session, then started the test as soon as they felt familiar with the system setup and the experimental procedure.

\begin{table}[t]
	\centering
	\caption{Rating scheme for subjective evaluation.}\vspace{-0.1 in}
	\begin{tabular}{|c||c|} \hline
		Rating level & Description \\  \hline
		5 & no difference to the undisturbed signal (perfect) \\  \hline
		4 & slightly perceptible disturbing (high quality) \\  \hline
		3 & disturbed (low quality) \\  \hline
		2 & strongly disturbed (very low quality) \\  \hline
		1 & completely distorted (unacceptable) \\  \hline
	\end{tabular}
	\label{Tab::subjectiveRating} \vspace{-0.1 in}
\end{table}

\begin{figure}[t]
	\centering
    \includegraphics[height=2.2cm,width=0.35\textwidth]{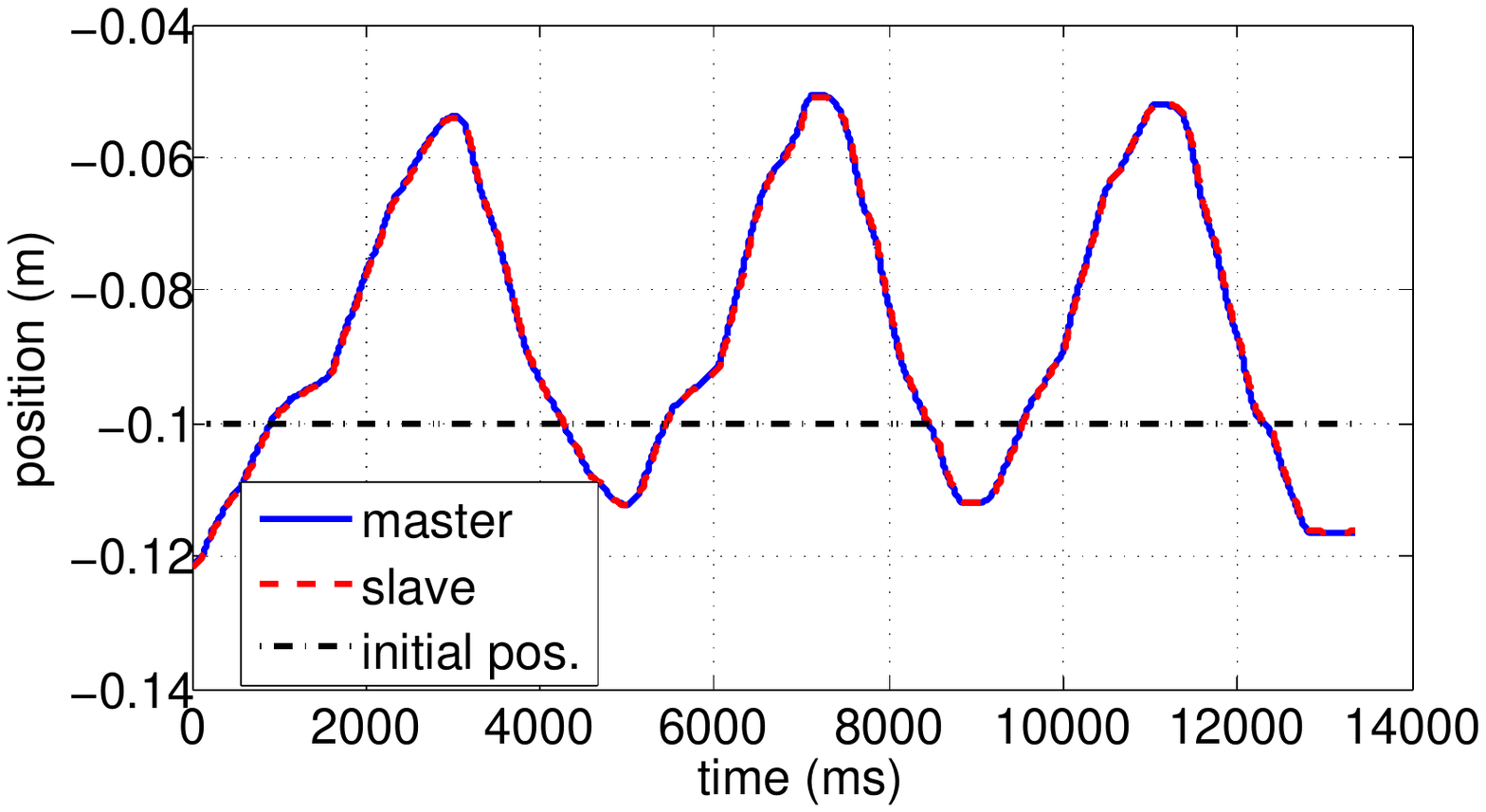}  \\
	\footnotesize (a) The master and slave position signals for 10 ms delay.
    \vspace{0.01in}
	\vfil
	\includegraphics[height=2.2cm,width=0.35\textwidth]{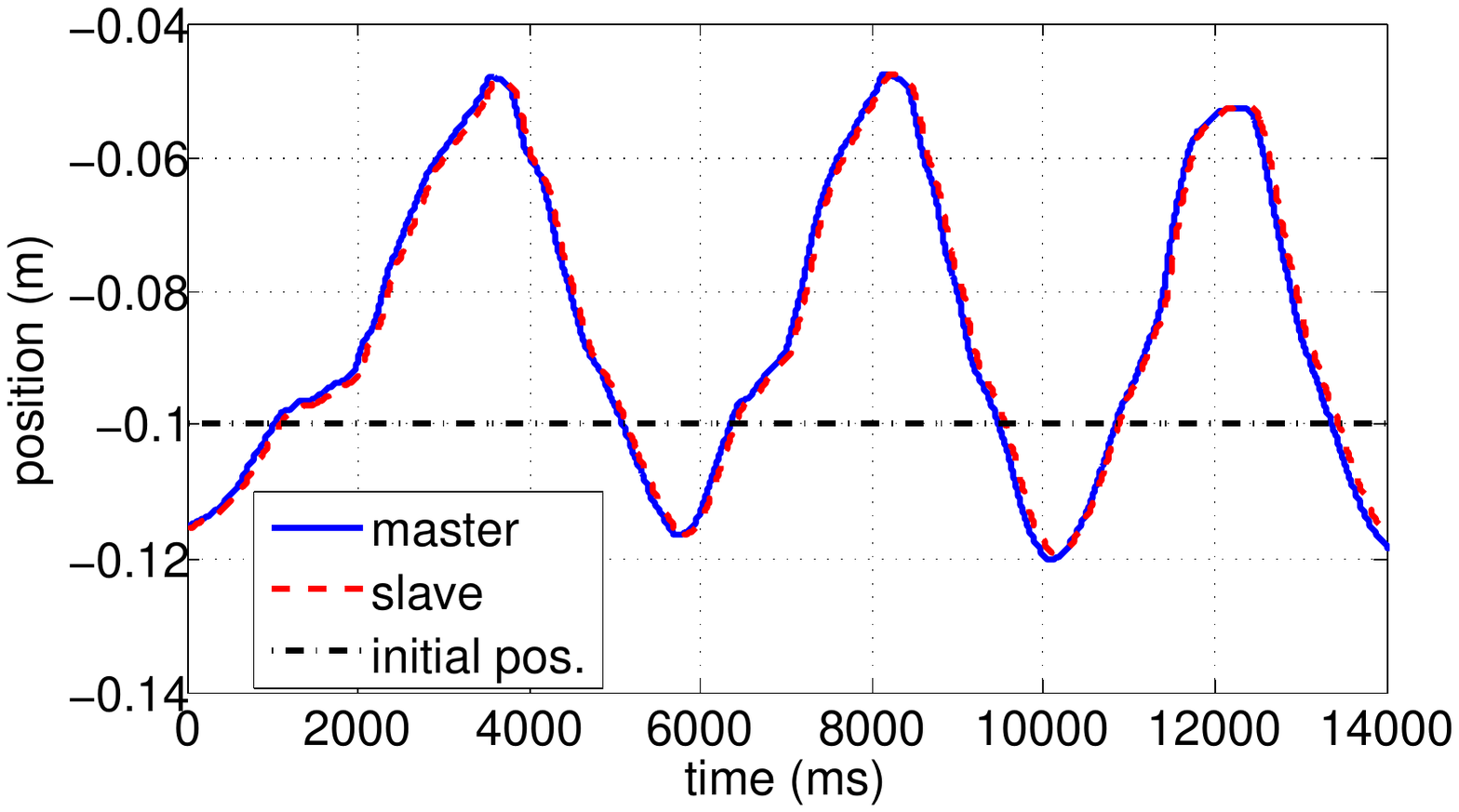}	 \\		 	
    \footnotesize (b) The master and slave position signals for 100 ms delay.
    \vspace{0.01in}
	\vfil
	\includegraphics[height=2.2cm,width=0.35\textwidth]{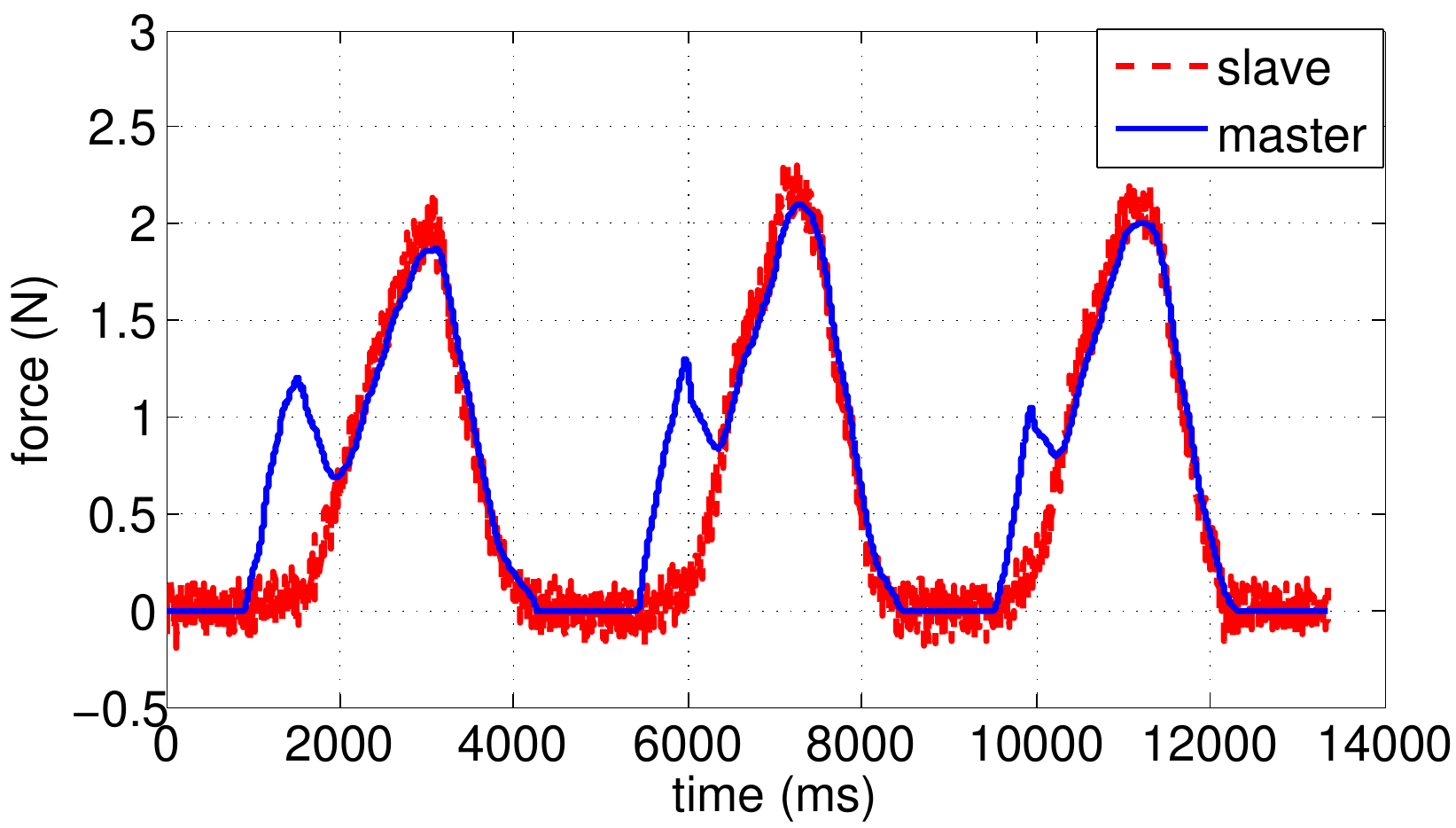}	 \\
	\footnotesize (c) The master and slave force signals for 10 ms delay.
    \vspace{0.01in}
	\vfil
	\includegraphics[height=2.2cm,width=0.35\textwidth]{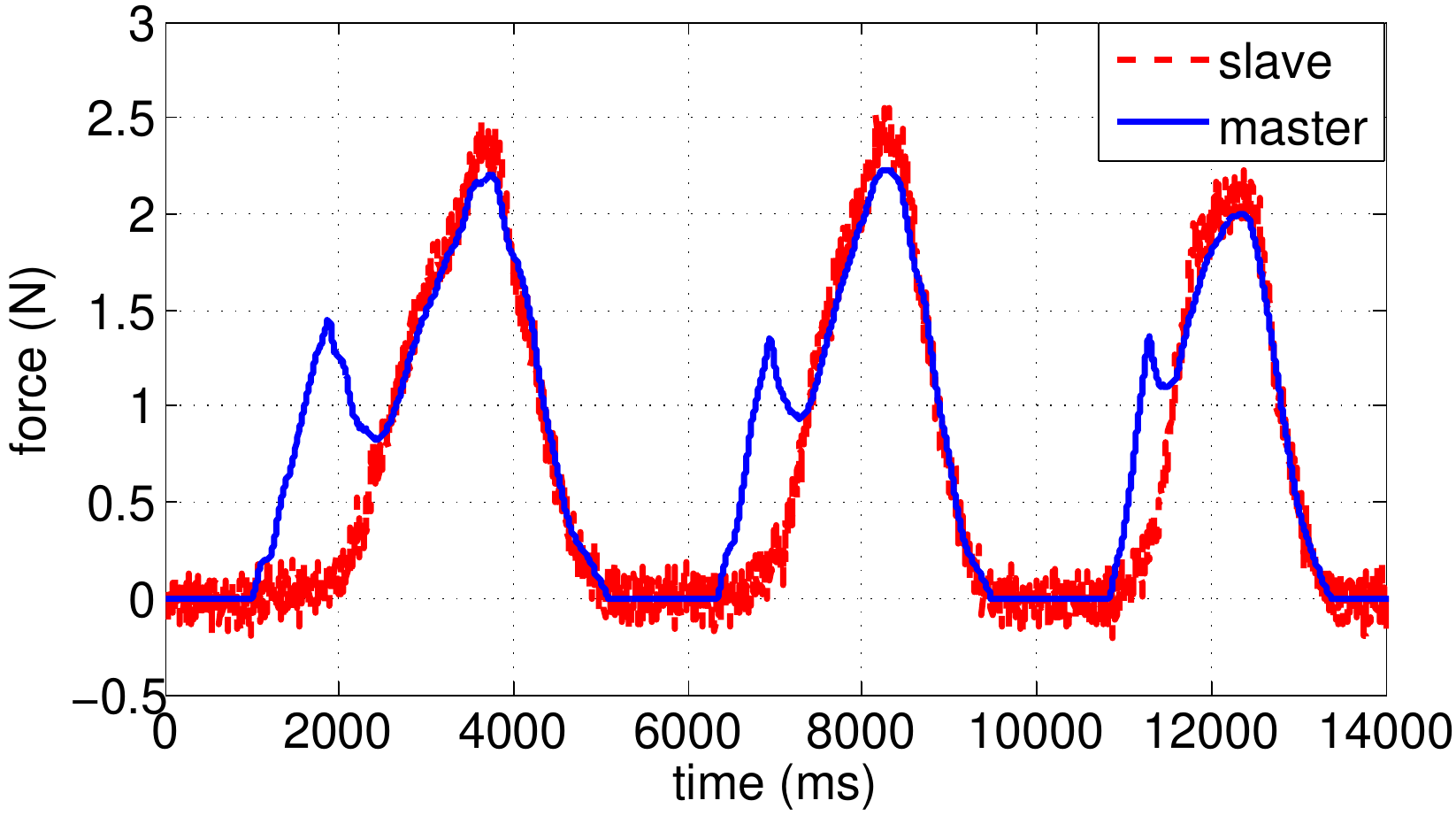} \\		 	\footnotesize (d) The master and slave force signals for 100 ms delay.
    \vspace{0.01in}
	\vfil
	\includegraphics[height=2.3cm,width=0.35\textwidth]{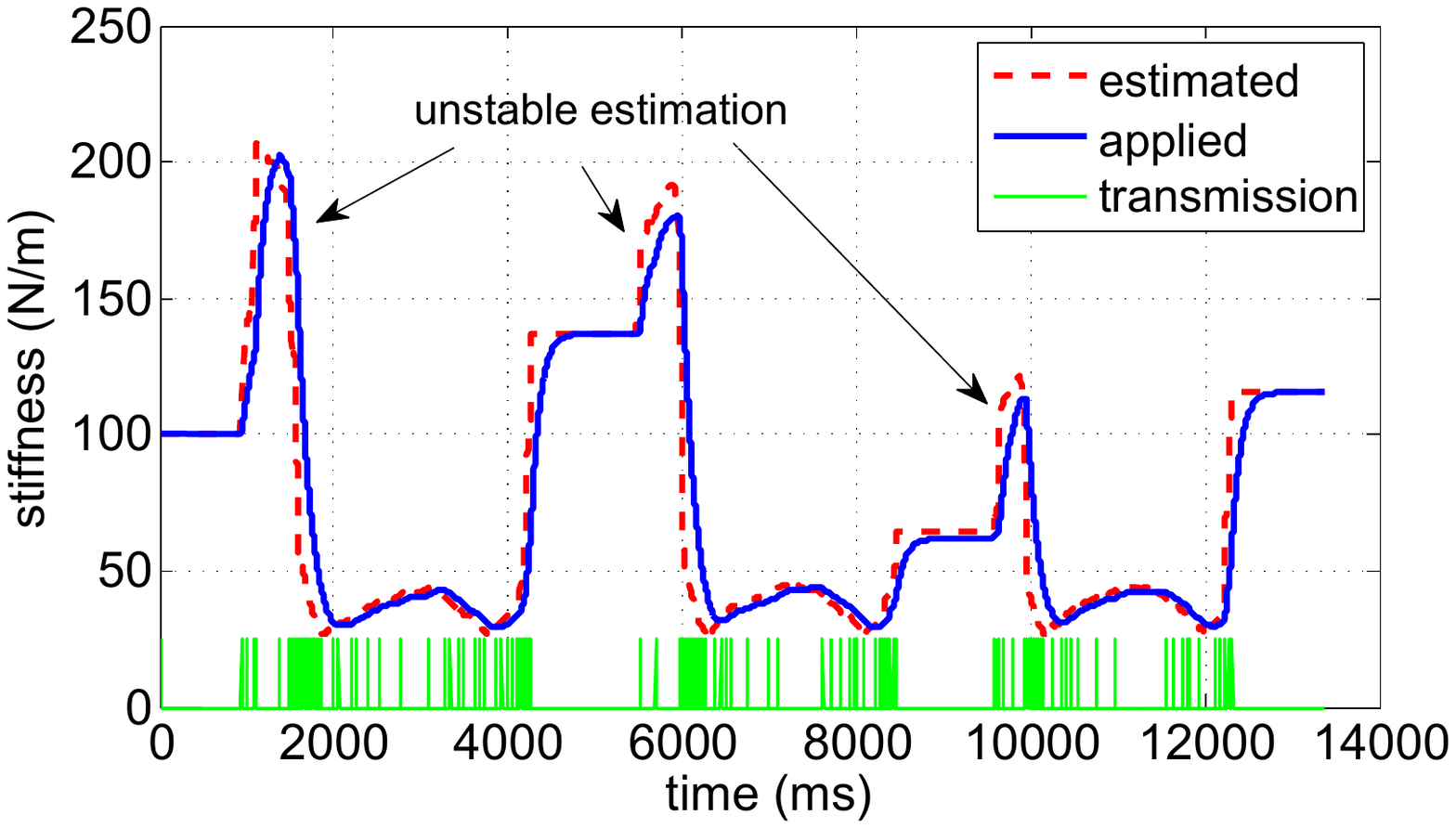}   \\
	\footnotesize (e) Estimated, transmitted, and applied stiffness values for 10 ms delay.
	\vfil
	\includegraphics[height=2.3cm,width=0.35\textwidth]{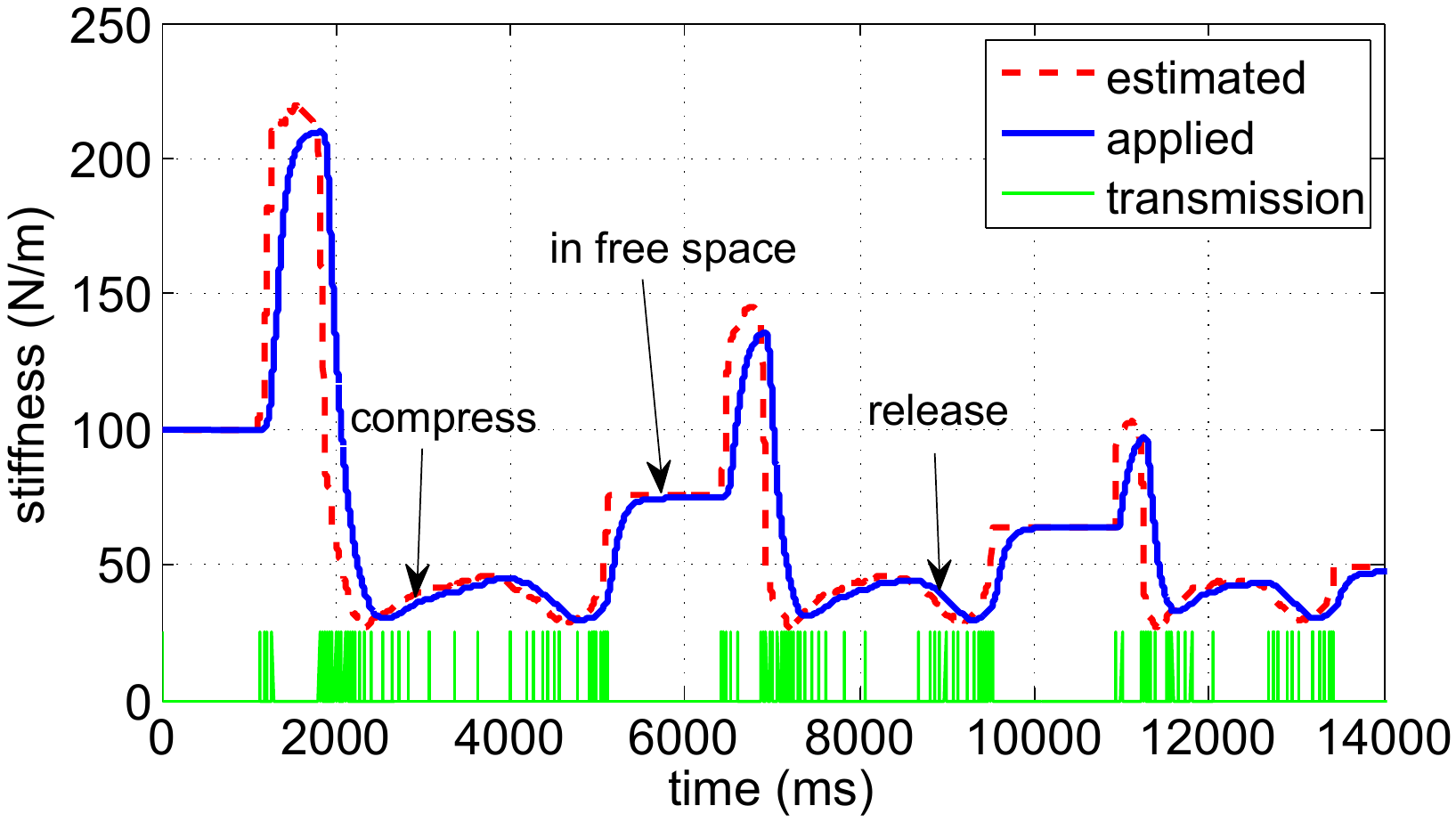} \\
	\footnotesize (f)  Estimated, transmitted, and applied stiffness values for 100 ms delay.
\vspace{-0.1 in}
	\caption{Measurements of the position and force signals and the estimated stiffness for the MMT+PD method. }

	\label{ch5:Fig:modeling}
\end{figure}

\subsection{Environment Modeling for MMT+PD}

Before discussing the subjective evaluation, it is necessary to pay attention to the modeling results of the MMT+PD method. Different from the TDPA+PD method, in which the communication delay has a dominant influence on the system performance, the modeling accuracy is the key factor that affects the system performance of the MMT+PD method. This means that once the model of the MMT+PD method is fixed for a static or slowly varying environment, the teleoperation quality degrades only slowly with increasing delay.

As an example, the measurements of the position, force, and estimated stiffness for delays of 10 ms and 100 ms are shown in Fig.~\ref{ch5:Fig:modeling}. For both delays, similar master position inputs lead to similar force signals and parameter estimates. This verifies that the communication delay in the tested range has only minor effects on the system performance.

\begin{figure}[t]
	\centering
	\includegraphics[height=2.2cm,width=0.35\textwidth]{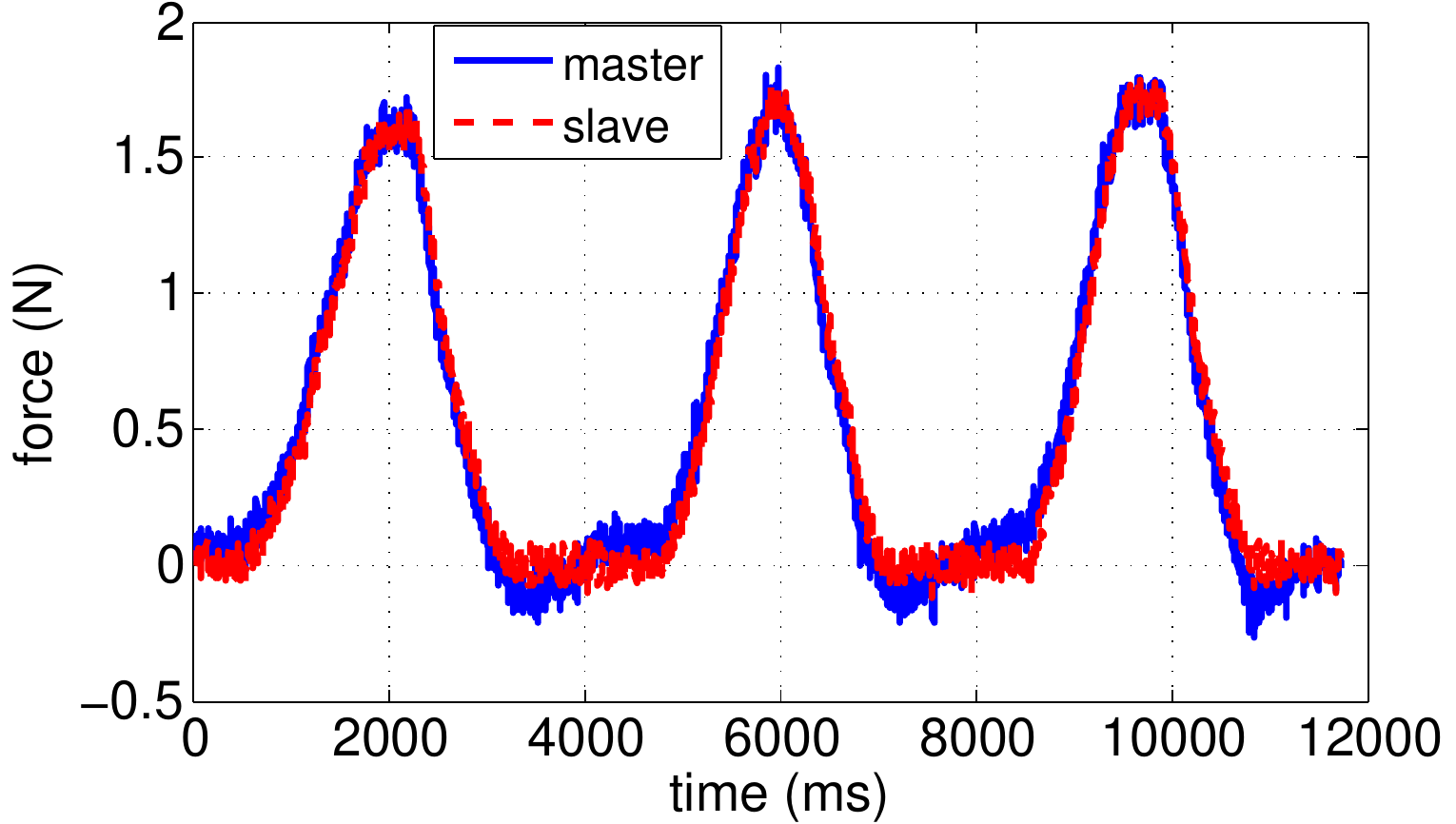} \\
	\footnotesize (a) Delay of 10 ms.
    \vspace{0.05in}
	\vfil
	\includegraphics[height=2.2cm,width=0.35\textwidth]{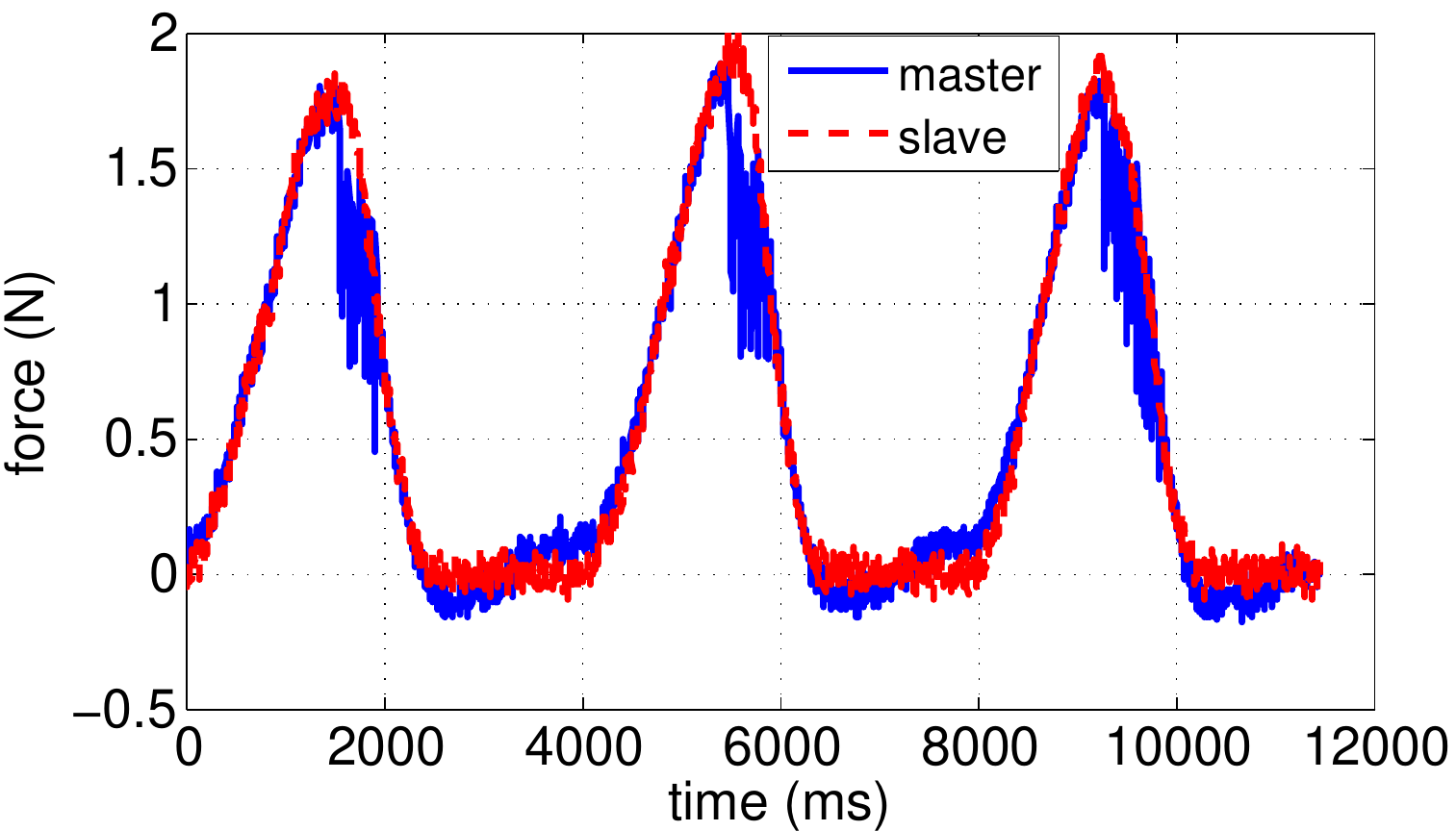} \\
    \footnotesize (b) Delay of 100 ms.
    \vspace{-0.05in}			
	\caption{Measurements of the force signals for the TDPA+PD method. }
\label{ch5:Fig:TDPAforce} \vspace{-0.0 in}
\end{figure}

In contrast, for the TDPA+PD method, a significant difference between the force signals for 10 ms delay and the force signals for 100 ms delay can be observed  from Fig.~\ref{ch5:Fig:TDPAforce}. This confirms that the communication delay has a higher influence on the performance of the TDPA+PD method than of  the MMT+PD method.

From Fig.~\ref{ch5:Fig:modeling} (c) and (d), we can observe unexpected peaks in the master force signals. This is caused by the overshooting in the stiffness estimation (unstable estimation) at every initial contact instants. After the estimates converge to the correct values, the master force, which is computed locally based on the applied linear spring model, follows the slave force without much deviation.

The estimated, transmitted, and applied stiffness values for 10 ms and 100 ms delays are shown in Fig.~\ref{ch5:Fig:modeling} (e) and (f). The use of the perceptual deadband-based (PD) data reduction approach avoids excessive transmission of the estimated stiffness data (see the green bars). The initial stiffness value for the local model is set to be 100 N/m before the slave's first contact with the object. Except for the time periods of overshooting and free space motion, the estimates slightly vary during the compress and release phases, leading to frequent packet transmission. Since a linear spring model is used to approximate the non-linear soft object, the estimated stiffness cannot be a constant value during the interaction. The more the object is compressed, the higher is the estimated stiffness. In addition, the passivity-based model update scheme proposed in \cite{Xu2014_TIM} is employed in this paper to guarantee stable and smooth changes in the applied stiffness values on the master side (represented by the blue solid lines).

Note that the strongly varying estimated stiffness at each initial contact leads to a mismatch between the master and slave force. This can disturb the subject's perception of the object stiffness and jeopardizes the teleoperation quality.

\subsection{Packet Rate}

Packet rate reduction over the network without introducing significant distortion is an important system capability to adaptively deal with different network conditions. It can be achieved by using a proper deadband parameter (DBP) for the TDPA+PD method as discussed in \cite{Xu2016_IEEEToH}. Furthermore, the MMT+PD method does not require a high update rate, especially for static or slowly varying environments. Model parameters are only updated when a significant model mismatch is detected. For the MMT+PD method, the stiffness is estimated every 1 ms based on the most recent 100 samples (position and force). The DBP in this experiment is set to 0.1 for both control schemes, indicating that a packet transmission is triggered when the change in force or estimated stiffness value is larger than 10\%.

\begin{figure}[t]
	\centering
	\includegraphics[width=0.35\textwidth]{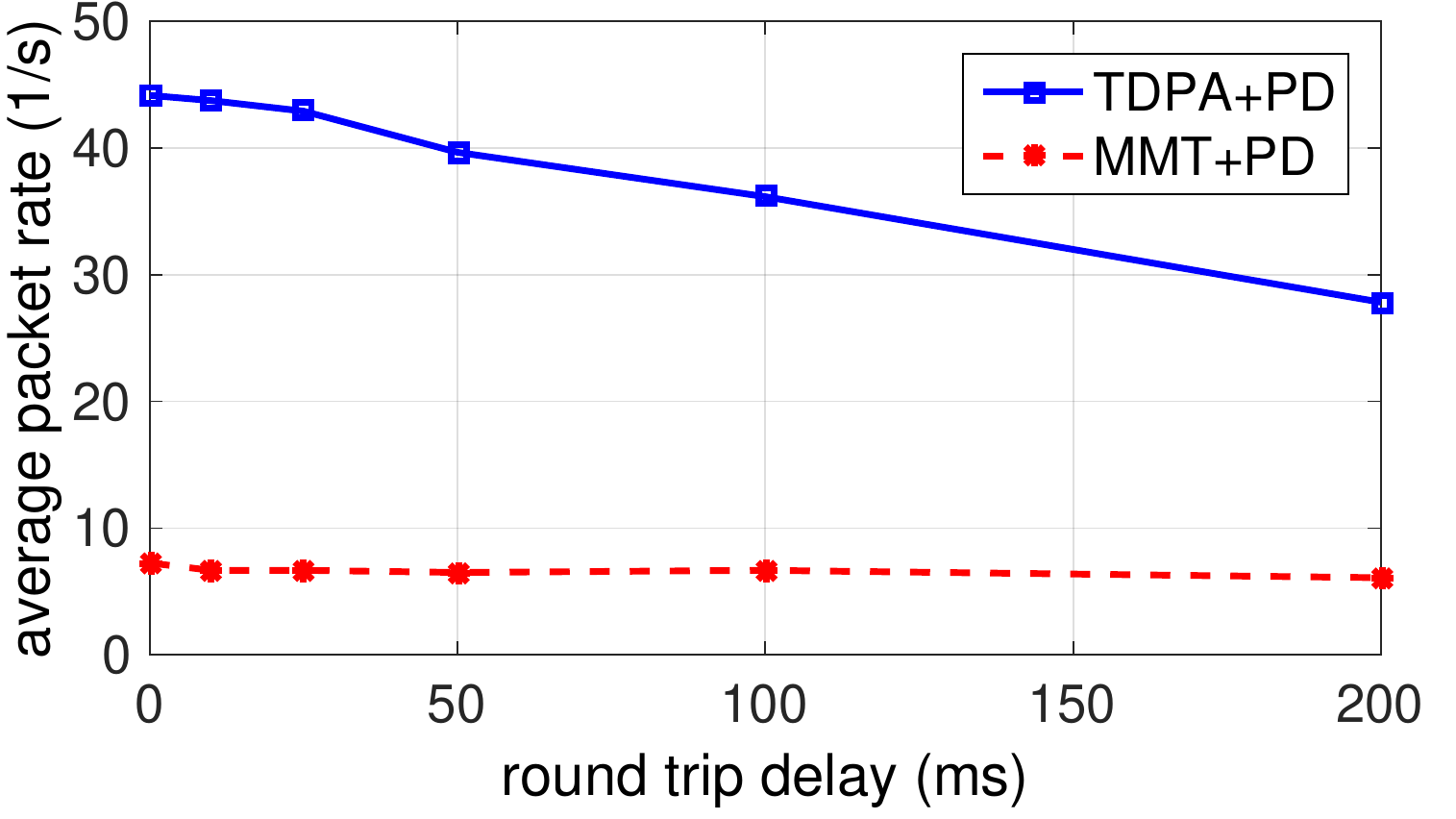}
\vspace{-0.1 in}
	\caption{Average packet rate over all subjects during the interaction with the soft object.}
    \vspace{-0.1in}
	\label{ch5:Fig:packRate} \vspace{-0.1 in}
\end{figure}

The packet rates for the two control schemes averaged over all subjects during their interaction with the soft object are shown in Fig.~\ref{ch5:Fig:packRate}. Although the applied local model mismatches the environment model, the average packet rates of the MMT+PD method are still much lower than for the TDPA+PD method. This is one of the strengths of the MMT+PD method compared to the TDPA+PD method.
For the MMT+PD method, triggering of packet transmission is concentrated in the periods of unstable estimation (e.g. at every initial contact and during the release). In addition, if the applied local model for the MMT method is more accurate, the packet rate during the interaction can be additionally reduced.

For the TDPA+PD method, the average packet rates over the tested delay range are 30$-$50 packets/s. It is noted that the packet rate of the TDPA+PD method is highly influenced by the interaction frequency. A larger interaction frequency will lead to more quickly varying velocity and force signals, resulting in higher packet rate.  In this experiment, the subjects controlled their interaction frequency to be lower than 1 Hz with the help of a visual guide.

\vspace{-0.1 in}
\subsection{QoE vs. Communication Delay} \label{subsec::QoE}

\begin{figure}[t]
	\centering
    \includegraphics[width=0.35\textwidth]{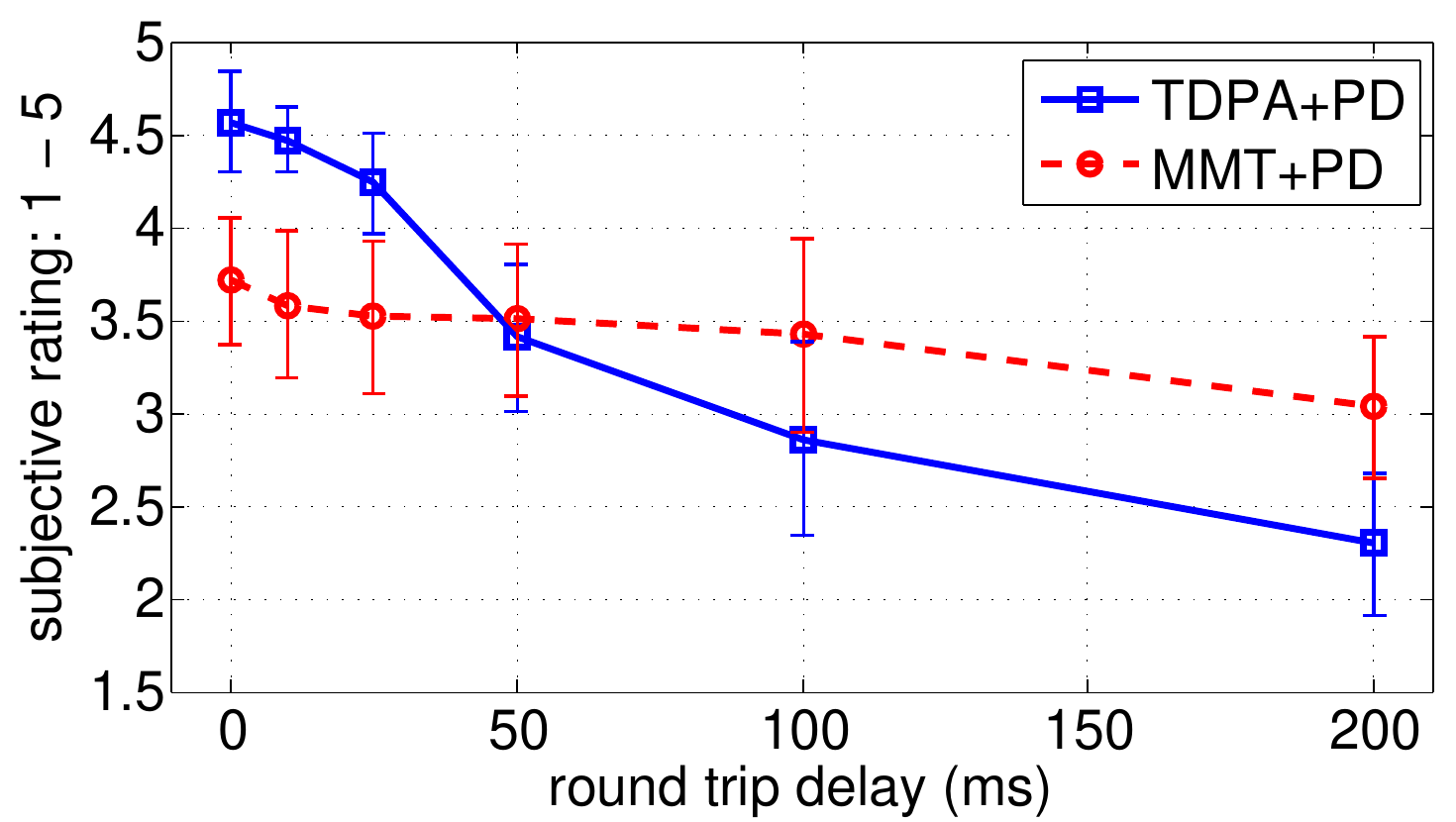}		\vspace{-0.1 in}
	\caption{Subjective ratings vs. communication delay for the two control schemes.} \vspace{-0.1in}
	\label{Fig::expRes}
\end{figure}

A quantitative evaluation of the subjective ratings (QoE) for the two control methods is illustrated in Fig.~\ref{Fig::expRes}. The QoE for both control methods degrades with increasing communication delay. For the MMT+PD method, the QoE is fairly stable, which confirms that the QoE of the MMT+PD approach mainly depends on the model accuracy, rather than the communication delay. In contrast, the TDPA+PD method is more sensitive to delay, as discussed in Section II.

According to Fig.~\ref{Fig::expRes}, the TDPA+PD method is able to provide relatively higher QoE than the MMT+PD method when the communication delay is small. However, the QoE of the TDPA+PD approach decreases quickly with increasing delay. This is because the subjects will perceive more vibrations and force jumps, as well as softer environment impedance when the delay grows. The overall rating results show that the subjects prefer the TDPA+PD method for small delays, and the MMT+PD method for large delays.

\begin{figure}[t]
  \centering
    \includegraphics[width=0.34\textwidth]{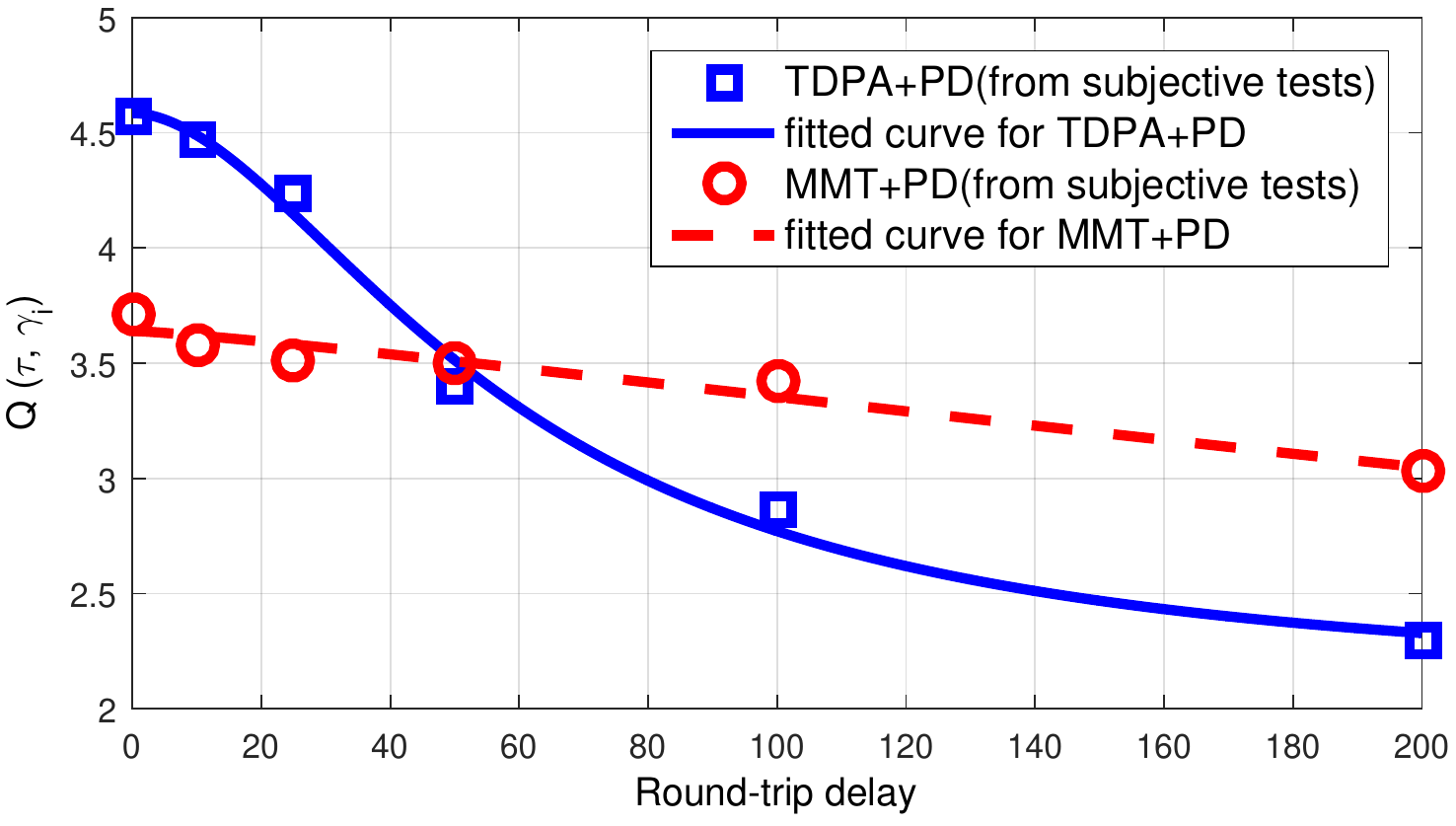} \vspace{-0.1 in}
\caption{QoE performance of different teleoperation systems with respect to delay (obtained via curve fitting).} \vspace{-0.1in}
 \label{Fig::qoeFitting}
\end{figure}

Based on the four-parameter logistic (4PL) \cite{campbell1994methods} curve fitting algorithm, we can obtain the QoE performance function with respect to the round-trip delay for both TDPA+PD and MMT+PD methods as:

\begin{equation}
Q(\tau,\gamma_i)=\frac{D_{\gamma_i}+(A_{\gamma_i}-D_{\gamma_i})}{1+(\frac{\tau}{C_{\gamma_i}})^{B_{\gamma_i}}}
\end{equation}
where $\gamma_1$ and $\gamma_2$ denote the TDPA+PD and MMT+PD methods, respectively. Thus, $A_{\gamma_1}=2.088, B_{\gamma_1}=-1.82, C_{\gamma_1}= 58.48, D_{\gamma_1}=4.585$, and $A_{\gamma_2}=0, B_{\gamma_2}=-1.187, C_{\gamma_2}= 793.7, D_{\gamma_2}=3.64$. An illustration of the curve fitted QoE performance is presented in Fig. \ref{Fig::qoeFitting}

By combing Eqn. (1) and (3), we can implement the proposed control scheme switching appraoch for this exemplary teleoperation use case. With a critical switching point of 50 ms (observed from Figs. \ref{Fig::expRes}-\ref{Fig::qoeFitting}), the TDPA+PD method should be adopted for short communication delays, otherwise, the MMT+PD method.

\section{CONCLUSIONS}

In this paper, we proposed a novel QoE-driven control scheme switching strategy for teleoperation systems. It is shown that the dynamic switching among different control schemes is essential in order to achieve the best QoE performance under different network conditions. We validated the feasibility of the proposed design with a dedicated case study. The simulation results confirm the efficiency of the proposed approach.

More importantly, this research revealed the intrinsic relationship among human perception, communication and control for time-delayed teleoperation systems. Therefore, we believe this paper can be considered as a fundamental reference for the future haptic communication research, especially in the area of joint optimization of communication and control.

\addtolength{\textheight}{-12cm}   

\bibliographystyle{IEEEtran}
\bibliography{ref}

\end{document}